\documentclass[preprint2]{aastex}

\newcommand{\footnt}[1]{\footnote{$^)$~#1}$^{)}$}
\newcommand{\codespc}{\vspace*{7pt}}

\slugcomment{~}

\shorttitle{Development of 2MASS Kit}
\shortauthors{Yamauchi, C.}

\begin{document}

%\title{Development of 2MASS Catalog Server Kit ---\\
%applying $x$$y$$z$-coordinate system to a huge catalog database}

\title{Development of 2MASS Catalog Server Kit}

\author{Chisato Yamauchi}
\affil{
 Astronomy Data Center,
 National Astronomical Observatory of Japan,
 2-21-1 Osawa, Mitaka, Tokyo, 181-8588, Japan
}
\email{chisato.yamauchi@nao.ac.jp}

\label{firstpage}

\begin{abstract}
We develop a software kit called ``2MASS Catalog Server Kit'' to easily construct
a high-performance database server for the 2MASS Point Source Catalog
(includes 470,992,970 objects) and several all-sky catalogs.
Users can perform fast radial search and rectangular search using
provided stored functions in SQL similar to SDSS SkyServer.
Our software kit utilizes open-source RDBMS,
and therefore any astronomers and developers can install our kit on
their personal computers for research, observation, etc.

Out kit is tuned for optimal coordinate search performance.
We implement an effective radial search using an orthogonal coordinate
system, which does not need any techniques that depend on HTM or
HEALpix.
Applying the $x$$y$$z$ coordinate system to the database index, we can
easily implement a system of fast radial search for relatively small 
(less than several million rows) catalogs.
To enable high-speed search of huge catalogs on RDBMS,
we apply three additional techniques: table partitioning, composite
expression index, and optimization in stored functions.
As a result, we obtain satisfactory performance of radial search for the
2MASS catalog.
Our system can also perform fast rectangular search.
It is implemented using techniques similar to those applied for radial
search.

Our way of implementation enables a compact system and will give
important hints for a low-cost development of other huge catalog
databases.
\end{abstract}

\keywords{Data Analysis and Techniques}

%\begin{keywords}
%astronomical data bases: miscellaneous;
%methods: miscellaneous
%\end{keywords}

\section{Introduction}
\label{section:intro}

Using huge object catalogs is more common in astronomical studies and
observations.
To support searching such catalogs, a variety of Web-based database
services have been developed.
For example,
NASA/IPAC Infrared Science Archive \citep[IRSA;][]{ber00},
VizieR \citep{och00} and
Virtual Observatory \citep[VO;][]{sza01} portal sites
are widely used in astronomical communities.
Some project teams distribute software to be installed in personal
computers to use such services via a network.
Astronomers and developers can use various catalogs with Web browsers
and such client-side software.

On the other hand,
there is software to search catalogs in offline environments.
For example, ``scat'' in the WCSTools package \citep{min06} is widely used in 
astronomical communities.
Such software is important for observatories with unstable or
narrowband networks or for personal studies that require huge catalog
entries.
Although SkyServer \citep{tha04} of Sloan Digital Sky Survey
\citep[SDSS;][]{yor00}
shows powerful flexibility of the programming interfaces based on
Structured Query Language (SQL) for catalog search,
it is not easy for end users to have such a high-performance search server 
in offline environments.

Therefore, we develop a software kit that enables any users to construct
a database system based on a relational data base management system
(RDBMS) in their personal computers and to quickly search a huge catalog
with functions similar to SDSS SkyServer.
The first target of our kit is the Two Micron All Sky Survey \citep{skr06}
Point Source Catalog (2MASS PSC),
which is huge but frequently used in astronomy.
We implement powerful functions into the kit using our various
techniques. 
One of the features of our techniques is applying an $x$$y$$z$ coordinate
system for fast radial search of a huge catalog.
This implementation might be rare and interesting for developers, and
we mainly report it in this article.

Our software kit is built on publicly available software.
In contrast, commercial RDBMS products have been used 
to develop previous Web-based database services of huge catalogs such as
SDSS SkyServer \citep{tha04}, 
WFCAM Science Archive \citep{ham07}, etc.
This article will also demonstrate the true power of an open-source RDBMS.

This article is organized as follows:
First, our software kit is introduced in \S\ref{overview}.
We show our overview of software design in \S\ref{overview_sd}.
In \S\ref{radial_search}, we explain details of our techniques and
reasonable implementation for high-speed radial search of our software
kit.
We additionally report techniques for rectangular search in 
\S\ref{rectangular}.
Summary is given in \S\ref{summary}.

Note that primitive data types of RDBMS are displayed in the following
way throughout this article:
{\tt INT2} is a signed two-byte integer, 
{\tt INT4} is a signed four-byte integer,
and 
{\tt FLOAT8} is a double-precision floating-point number.

%\begin{itemize}
% \item
%      {\tt INT2} ... signed two-byte integer
% \item
%      {\tt INT4} ... signed four-byte integer
%% \item
%%      {\tt INT8} ... signed eight-byte integer
% \item
%      {\tt FLOAT8} ... double-precision floating-point number
%\end{itemize}

\section{2MASS Catalog Server Kit}
\label{overview}

\begin{figure}[!t]
 %\epsscale{1.0}
 %\plotone{sshot_kit.eps}
 \plotone{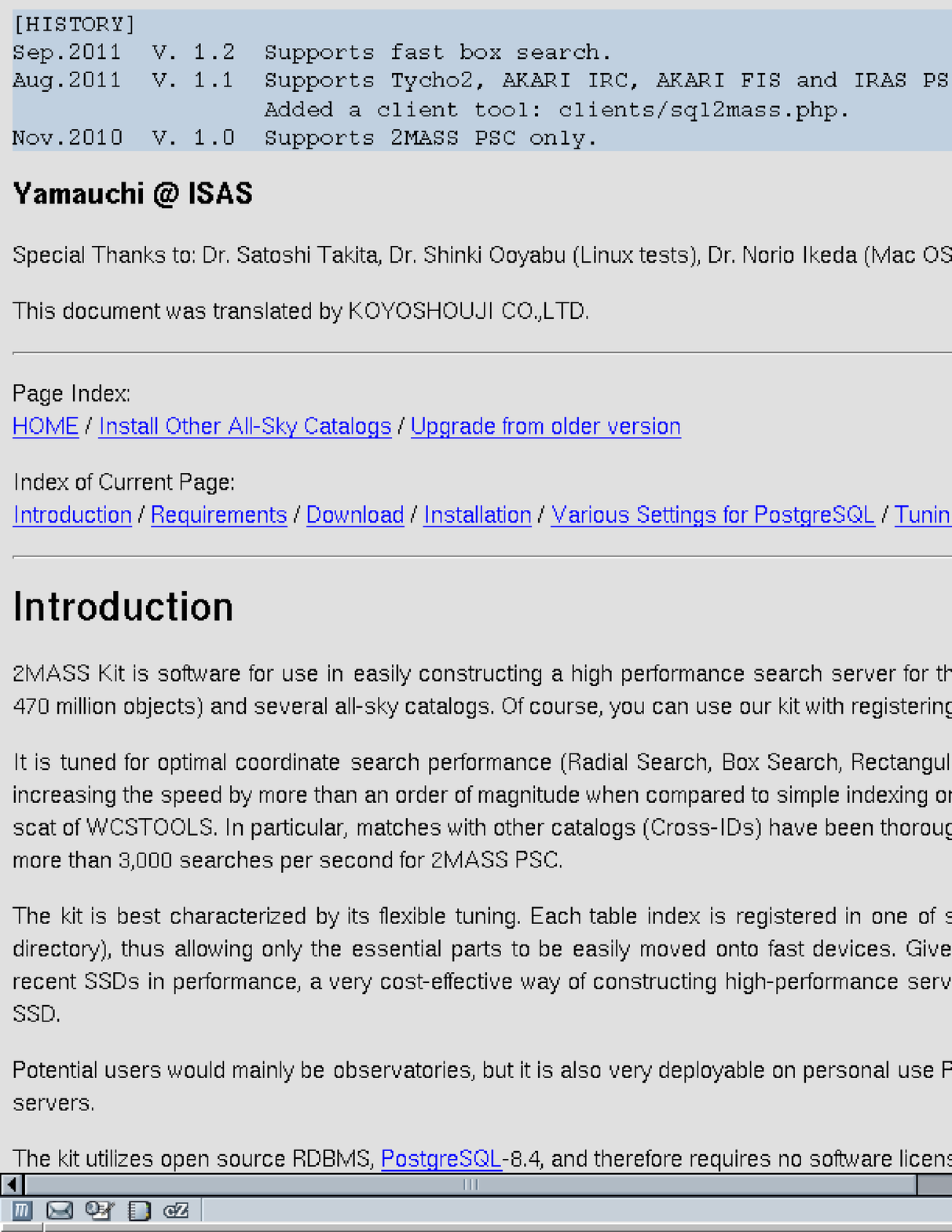}
 \caption{Web page of the 2MASS Catalog Server Kit.
 Users can easily install, use, and tune their own catalog
 database.
 This page includes enough contents for RDBMS beginners.
 }\label{fig:sshot_kit}
\end{figure}

The 2MASS Catalog Server Kit%
\footnt{http://www.ir.isas.jaxa.jp/$^\sim$cyamauch/2masskit/.
Software package and data set are available in this page.
We are planning to support USNO-B1.0, PPMXL, GSC-2.3, etc.
} 
(2MASS Kit) is a software package to
construct a high-performance search server of 2MASS PSC
and several all-sky catalogs 
on Linux, MacOSX, Solaris, and other UNIX systems.
To install this kit,
it is enough to prepare a standard personal computer with a hard drive
of 600 Gbyte or more.

The kit contains a complete data set of tables for 2MASS PSC,
SQL statements and sources of C language with a Makefile.
The HTML document (Fig. \ref{fig:sshot_kit})
shows step-by-step instructions for installation,
a tutorial for database beginners, reference of stored functions, and
several hints for tuning.
The instructions include a procedure for %installing the package,
setting up RDBMS and configuration of the operating system;
therefore, any users can easily construct their own catalog database
servers.
Users can search catalogs not only with flexible SQL but also with
several useful stored functions prepared by our kit.
Using the functions in users' SQL statements,
users can perform fast radial and rectangular search with very small SQL
statements with coordinate conversions (e.g., J2000 to Galactic).
Of course, users do not have to know algorithm and indexing about
typical searches.
We show an example SQL statement of a radial search of Galactic
coordinate (0,0) with 0.2$'$ radius:\codespc\\
\verb|SELECT fJ2L(o.ra,o.dec) as l,|\\
\verb|       fJ2B(o.ra,o.dec) as b,|\\
\verb|       o.j_m,o.h_m,o.k_m, n.distance|\\
\verb|FROM fTwomassGetNearbyObjEq(|\\
\verb|           fG2Ra(0,0),fG2Dec(0,0),0.2) n,|\\
\verb|     twomass o|\\
\verb|WHERE n.objid = o.objid;|\codespc\\
where {\tt fJ2L()} and {\tt fJ2B()} convert J2000 to Galactic, 
{\tt fG2Ra()} and {\tt fG2Dec()} convert Galactic to J2000,
and {\tt fTwomassGetNearbyObjEq()} is a stored function that performs
fast radial search of 2MASS PSC with an optimized algorithm.

The kit is also characterized by its flexible tuning. Each table and
index for 2MASS PSC is registered in one of seven table spaces (each
resides in a
separate directory), thus allowing only the essential parts to be easily
moved onto fast devices. Given the terrific evolution that has taken
place with recent solid-state drives (SSDs) in performance, a very
cost-effective way of constructing high-performance servers is moving
part, or all, of the table indices to a fast SSD.

Before installing the 2MASS Kit,
users can confirm the performance of our kit using the SQL Search Tool%
\footnt{http://darts.jaxa.jp/ir/akari/cas/tools/search/sql.html.}
of the AKARI Catalogue Archive Server \citep[AKARI-CAS;][]{yam11}.
AKARI-CAS is developed to search AKARI All-Sky Catalogues
based on imaging data obtained by 
the Far-Infrared Surveyor \citep[FIS;][]{kaw07} and
the Infrared Camera \citep[IRC;][]{ona07}
built on AKARI satellite,
and it also supports fast search for 2MASS PSC.
To perform this search,
AKARI-CAS has the source codes of 2MASS Kit.
Of course, our kit utilizes open-source RDBMS, PostgreSQL-8.4, and
therefore requires no software licensing fees.

\section{Overview of Software Design}
\label{overview_sd}

Choosing an appropriate RDBMS product is important for our software
design, since functions to support various users' requirements depend
on RDBMS products.

Users will search the 2MASS PSC using various criteria. 
Therefore, the RDBMS product should meet recent SQL standards and must have
enough search performance.
In addition, some users might not have knowledge about indices of tables.
To support such users,
the kit has to provide some functions to minimize SQL statements for
typical searches in astronomy.
Therefore, the RDBMS product should have high coding flexibility of stored
functions.
We investigated some open-source RDBMS products when developing
AKARI-CAS.
We found that PostgreSQL perfectly satisfies the preceding requirements.
See also \citet{yam11} for an investigation about RDBMS products.

Fast positional search is indispensable for astronomy,
even if the database has huge catalogs.
Fortunately, PostgreSQL has some special features to handle huge tables.
For example, PostgreSQL supports table inheritance that is useful for
table partitioning and has a ``constraint-exclusion'' feature to
allow us a seamless access to the partitioned tables.
To obtain the best performance of positional search, i.e., radial search
and rectangular search,
we apply such features and write our codes in stored functions for more
optimization.

We mainly tune the performance of radial search, since it is most
important for astronomical catalog databases.
Our severe test for it is done by cross-identification as multiple
radial searches.
It is the main theme of this article.

One of the advantages of PostgreSQL is having many built-in functions usable
in SQL statements.
Together with them, 
coordinate conversions shown in \S\ref{overview} such as J2000-to-Galactic conversion are supported by newly created stored functions that
contain some codes of 
wcstools-3.8.%
\footnt{http://tdc-www.harvard.edu/wcstools/.}
Our kit provides further functions that are also available in AKARI-CAS.
See \citet{yam11} for general technical know-how for creating stored
functions in PostgreSQL.
%See also our software kit page and source codes for details.

\section{Technical Design of Radial Search}
\label{radial_search}

\subsection{Basic Algorithm}
\label{basic_method}

The radial search is the most typical query in the catalog database
services.
However, this search using RDBMS cannot be optimized in a
straightforward way, because an index of RDBMS is useful for the case:
\begin{equation}
 S_{\rm min} \le f(c_{A_n},c_{B_n}, ... ,c_{X_n}) \le S_{\rm max},
\end{equation}
where $S_{\rm min}$ and $S_{\rm max}$ are the search criteria and
$c_{A_n}$,$c_{B_n}$,$...$ are data of columns A, B, etc.;
however, an index cannot be created for following case:
\begin{equation}
 S_{\rm min} \le f(c_{A_n},c_{B_n},S_1,S_2) \le S_{\rm max},
\label{eq:rad}
\end{equation}
where $S_1$ and $S_2$ are also the search criteria.
The radial search corresponds to the case (\ref{eq:rad}),
i.e., $f()$ is a function that takes a pair of positions and returns an
angular distance, and $S_{\rm max}$ is the search radius.

To enable fast radial search applying the database index,
some special methods based on spatial splitting have been devised, such
as Hierarchical Triangular Mesh%
\footnt{http://skyserver.org/HTM/.}
\citep[HTM;][]{kun00}
or
HEALPix%
\footnt{http://healpix.jpl.nasa.gov/.}
\citep{gor05}.
Their methods divide the sky into many areas, assign each area
the unique ID, and give each object a corresponding ID from which
the one-dimensional index is created.

\begin{figure}[!t]
 %\epsscale{1.0}
 %\plotone{idea.eps}
 \plotone{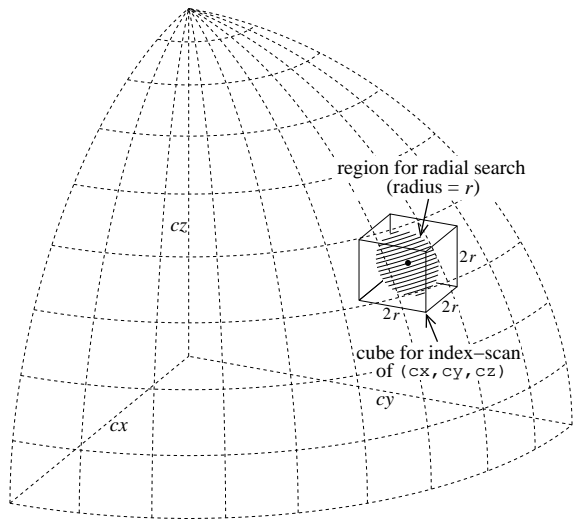}
 \caption{Concept of our radial search with search radius $= r$.
 Catch objects within a cube $2r$ each side using index scan on
 {\tt (cx,cy,cz)}, then select objects within the strict search area
 (striped pattern) from them.
 }\label{fig:concept}
\end{figure}

We do not use such techniques, but use the more simple and cost-effective
way. 
Figure \ref{fig:concept} shows the concept of our radial search.
The most important point of this concept is the use of the 
$x$$y$$z$ coordinate
for the database index.

In our databases, the object tables have columns of unit vectors 
({\tt cx}, {\tt cy}, {\tt cz}) presenting J2000 source positions.
We create a composite index on ({\tt cx}, {\tt cy}, {\tt cz})
and write stored functions to execute the following procedure:
\renewcommand{\labelenumi}{\arabic{enumi}.}
\begin{enumerate}
 \item
      Catch objects within a cube of the size 
      $2r$$\times$$2r$$\times$$2r$
      %each side 
      using index scan on ({\tt cx}, {\tt cy}, {\tt cz}).
 \item
      Select objects within the strict search circle on the celestial
      sphere from the result of step 1.
\end{enumerate}

The feature of our algorithm is that it requires almost no calculation
before executing the index scan, and the efficiency is quite high for a
small search radius.
In addition, we do not have to implement special processing for polar
singularity.
\citet{tan93} pointed out the advantage of applying orthogonal
coordinate system to avoid polar singularity.

\subsection{Implementation for Small Catalogs}
\label{for_small_catalogs}

In this section, we present our implementation and test results
of radial search of the 
AKARI/IRC Point Source Catalogue \citep[AKARI/IRC PSC;][]{ish10},
including 870,973 objects.
It is relatively small compared with the SDSS or 2MASS catalog;
therefore, we implement the system only with the basic method described
in \S\ref{basic_method}.

To enable fast radial search,
we construct our table and index with the following procedure:
\renewcommand{\labelenumi}{\arabic{enumi}.}
\begin{enumerate}
 \item
      Register all rows of all columns 
      of AKARI/IRC PSC into a table {\tt AkariIrc}.
      The table includes a primary key {\tt objID} of {\tt INT4} type,
      J2000 source position ({\tt ra}, {\tt dec}) of
      {\tt FLOAT8} type
      and unit vector ({\tt cx}, {\tt cy}, {\tt cz}) of {\tt FLOAT8}
      type converted from ({\tt ra}, {\tt dec}).
 \item
      Create a composite index
      on ({\tt cx}, {\tt cy}, {\tt cz})
      by the following SQL statement:%
      \footnt{In the case of PostgreSQL,
      we have to do {\tt VACUUM ANALYZE;} after creating indices.}%
\codespc\\
      \verb|CREATE INDEX akariirc_xyz |\\
      \verb|       ON AkariIrc(cx,cy,cz);|
\end{enumerate}

To perform a radial search, we create SQL stored functions.
For example, the source code of a stored function to obtain an 
{\tt objID} of the object whose distance from search position is the smallest
in the search region is given next:%
\footnt{See Appendix A for each stored function written in this definition.}%
\codespc\\
{\footnotesize
\verb|CREATE FUNCTION |\\
\verb|       fAkariIrcGetNearestObjIDEq(FLOAT8,FLOAT8,FLOAT8)|\\
\verb|RETURNS INT4 AS|\\
\verb|'SELECT o.objID|\\
\verb| FROM (|\\
\verb|  SELECT objID,|\\
\verb|    fDistanceArcMinXYZ(fEq2X($1,$2),fEq2Y($1,$2),|\\
\verb|               fEq2Z($1,$2),cx,cy,cz) as distance|\\
\verb|  FROM AkariIrc|\\
\verb|  WHERE |\\
\verb|  (cx BETWEEN fEq2X($1,$2) - fArcMin2Rad($3) AND|\\
\verb|              fEq2X($1,$2) + fArcMin2Rad($3)) AND|\\
\verb|  (cy BETWEEN fEq2Y($1,$2) - fArcMin2Rad($3) AND|\\
\verb|              fEq2Y($1,$2) + fArcMin2Rad($3)) AND|\\
\verb|  (cz BETWEEN fEq2Z($1,$2) - fArcMin2Rad($3) AND|\\
\verb|              fEq2Z($1,$2) + fArcMin2Rad($3))|\\
\verb| ) o|\\
\verb| WHERE o.distance <= $3|\\
\verb| ORDER BY o.distance|\\
\verb| LIMIT 1'|\\
\verb|IMMUTABLE LANGUAGE 'sql';|\codespc\\
}%
where {\tt \$1}, {\tt \$2}, and {\tt \$3} are the arguments of this
stored function; ({\tt \$1}, {\tt \$2}) is the center position in J2000
of the search region; and {\tt \$3} is the search radius.

If we cutout from {\tt SELECT o.objID} to 
{\tt WHERE o.distance <= \$3} in the preceding source and execute it
by giving actual values for
{\tt \$1}, {\tt \$2}, and {\tt \$3}, it performs a radial search.

\begin{table}[!t]
\begin{center}
\caption{Results of match-up of AKARI/FIS BSC with
AKARI/IRC PSC.
The hardware is a Core2Quad Q9650 (3.0GHz) CPU on
GIGABYTE GA-EX38-DS4 with 8 Gbyte DDR2-800 memory and
an Adaptec RAID 2405 with 1 Tbyte SATA2 HDD$\times$2 (RAID1).
}\label{table:result_fis_irc} {
\vspace*{6.0pt}
\begin{tabular}{lc}
\hline
\hline
Condition & Elapsed time \\
\hline
Just after OS rebooting & 42.0 s \\
Second trial            & 19.8 s \\
\hline
\end{tabular} 
}
\end{center}
\end{table}

To evaluate the performance of our radial search,
we try a cross-identification as multiple radial searches
using {\tt fAkariIrcGetNearestObjIDEq()}.
We show our test results of matching up all objects of
AKARI FIS Bright Source Catalogue%
\footnt{AKARI/FIS All-Sky Survey Bright Source Catalogue Version 1.0 Release Note (Yamamura et al. 2010), {\tt http://www.ir.isas.jaxa.jp/AKARI/Observation/PSC/Pu\\blic/RN/AKARI-FIS\_BSC\_V1\_RN.pdf}.}
(AKARI/FIS BSC; including 427,071 objects)
with all AKARI/IRC PSC objects within 0.25$'$ radius
in Table \ref{table:result_fis_irc}.
Here is the SQL statement for this test:\codespc\\
{\footnotesize
\verb|SELECT count(fAkariIrcGetNearestObjIDEq(ra,dec,0.25)) |\\
\verb|FROM AkariFis;|\codespc\\
}%
This returns 19,267 matches.

AKARI catalogs are small enough compared with memory capacities of
present computers.
Therefore, users generally have to wait only 30 s or so, even for
cross-identification. 
We can implement radial search of the catalogs including less than
several million objects with our simple method by applying
the $x$$y$$z$ coordinate.
See also \citet{yam11} for more applications for AKARI catalogs using
our techniques.

\subsection{Implementation for 2MASS PSC}
\label{implementation_2mass}

\subsubsection{Data Size Limit of Simple Radial Search Implementation}
\label{row_limit}

In \S\ref{for_small_catalogs},
we store all the contents of a catalog into a table,
create a composite index on ({\tt cx}, {\tt cy}, {\tt cz}),
and write a simple SQL stored function to perform a radial search.
However, there is a limit of row numbers for this simple
implementation described in \S\ref{for_small_catalogs}.
This limit is caused by two factors:
(1) enlargement of processing and
(2) the bottleneck of disk I/O access.
First, we show the behavior of factor 1 here.

We test the performance of cross-identification as
multiple radial searches. 
We use a subset of AKARI/IRC PSC and a subset of 2MASS PSC
for the cross-identification.
The table of the subset of 2MASS PSC includes all columns and additional
primary key {\tt objid} of {\tt INT4} and unit vector
({\tt cx}, {\tt cy}, {\tt cz}) of {\tt FLOAT8}
calculated at the data registration,
and a composite index on the unit vector is created.
To examine the performance dependency on the number of data entries, we
prepare five cases;
5099652, 10226706, 20294711, 40688903 and 82092729,
selected by declination range from the south pole.
Then we test the performance of cross-identification for each case.
The following SQL statement is an example of the test:\codespc\\
{\footnotesize
\verb|SELECT count(|\\
\verb|      fTwomassGetNearestObjIDEq(o.ra, o.dec, 0.25))|\\
\verb|FROM (SELECT * FROM AkariIrc|\\
\verb|      WHERE dec < -74.600006) o;|\codespc\\
}%
A criterion of `{\tt dec < -74.60006}' is to choose AKARI/IRC objects
within the region corresponding to the 2MASS PSC subset.

\begin{figure}[!t]
 %\epsscale{1.0}
 %\plotone{limit.eps}
 \plotone{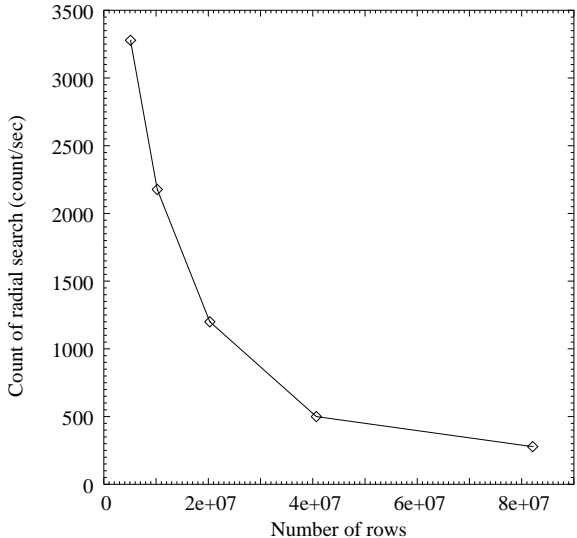}
 \caption{Performance test of simple radial search implementation
 applied for a huge catalog.
 We test cross-identification between a subset of AKARI/IRC PSC and subset
 of 2MASS PSC for different row sizes of the 2MASS PSC table;
 count of radial search per second is plotted against the number of
 rows in the 2MASS PSC table.
 Hardware is a $2\times$ Opteron2384 CPU on a Supermicro dual-processor
 board with 32 Gbyte memory.
 }\label{fig:limit}
\end{figure}

Figure \ref{fig:limit} shows the result of our tests.
In the case of the smallest data size about 5.1 million objects,
searches of more than 3000 counts/s are performed.
However, processing speed is rapidly dropped with increasing number of
rows.
Each measurement in Figure \ref{fig:limit} was obtained from the
median of the last three successive runs;
i.e., all measurements were taken under the condition of enough cached
data in the main memory.
Therefore, this result means that it is impossible to obtain acceptable
performance for severe search requirements with huge catalogs using this
simple implementation, even if there is no bottleneck of disk I/O
access.

\subsubsection{Strategy to Break the Data Size Limit}
\label{to_break_limit}

As shown in \S\ref{row_limit},
we cannot achieve sufficient performance under severe search
requirements of huge catalogs using simple implementation.
Moreover, we can easily expect that the bottleneck of disk I/O becomes
a serious problem for actual use.
To break such limits, we consider to optimize the design of table
relations, indices and stored functions.
Our strategy for implementation of huge catalogs is as follows;
\renewcommand{\labelenumi}{\arabic{enumi}.}
\begin{enumerate}
 \item
      Reduce the height of the nonunique index.
 \item
      Reduce the file size of the data set for
      performing a radial search.
 \item
      Note that unique indices (e.g., primary key) give enough
      performance for a huge table.
 \item
      Minimize CPU time for additional processing.
 \item
      Carry out experimental tuning.
\end{enumerate}

\subsubsection{Design of Rable Relation}
\label{design_table_relation}

Considering the strategy in \S\ref{to_break_limit},
we determine the design of table relation as follows:
\renewcommand{\labelenumi}{\arabic{enumi}.}
\begin{enumerate}
 \item
      We apply the table partitioning technique.
 \item
      We prepare a special table set consisting of only necessary
      columns for each search purpose.
 \item
      We store the object positions into integer ({\tt INT4}) columns
      in this special table set.
      These integer values are converted and scaled from the original
      floating-point values of right ascension and declination.
\end{enumerate}

\begin{figure}[!t]
 %\epsscale{1.0}
 %\plotone{relation.eps}
 \plotone{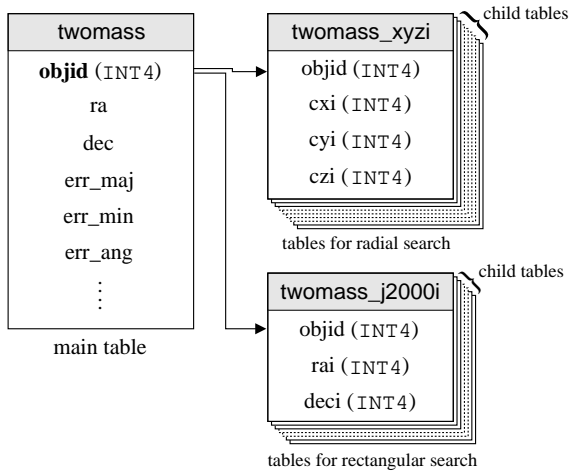}
 \caption{Design of table relation to enable fast positional search
 of 2MASS PSC.
 Columns ({\tt cxi}, {\tt cyi}, {\tt czi}) and
 ({\tt rai}, {\tt deci}) are the unit vectors, right ascension and
 declination, respectively.
 Their integer values are converted and scaled from floating-point values
 in columns ({\tt ra}, {\tt dec}) in the {\tt twomass} table.
 }\label{fig:relation}
\end{figure}

Figure \ref{fig:relation} shows our design of table relation.
The main table ``{\tt twomass}'' has 470,992,970 entries
(without partitioning%
\footnt{In our tests, 
partitioning of main table decreases the performance of join on the
primary key.}%
)
and is basically supposed to be searched with the primary key.
On the other hand, 
table partitioning is applied to the
table sets ``{\tt twomass\_xyzi}'' and ``{\tt twomass\_j2000i}'' to reduce the 
height of the nonunique index in each table.
These two table sets are optimized for radial search and rectangular
search (see \S\ref{rectangular}), respectively.

Table partitioning is supported via table inheritance in PostgreSQL.
A parent table has column definition and empty rows, and child tables
have the same columns as those of the parent and a number of rows. 
We determine the contents of child tables with the range partitioning
using values of declination.
This partitioning is implemented so that child tables have almost
the same number of rows.
The optimal number of partitions is discussed in \S\ref{partitioning}.

The values of ({\tt cxi}, {\tt cyi}, {\tt czi}) are converted from
the original right ascension and declination and scaled 
between $-2$$\times$$10^9$ and $2$$\times$$10^9$ (integer)
so that the spatial resolution is fine enough for astronomical object
catalogs.
When performing a radial search,
this integer version of the unit vector is restored into floating-point
values to calculate angular distance.

The data size of the table set is about 20 Gbyte,
which reduces disk read traffic and disk seek time.

\subsubsection{Index}

We can notice that spatial resolution of the composite index on 
({\tt cxi}, {\tt cyi}, {\tt czi})
does not have to be that of {\tt INT4}, 
since the index is only used to preselect objects within a cube.
Therefore, we can reduce the size of the index on
({\tt cxi}, {\tt cyi}, {\tt czi})
using the composite expression index so that the index is created in {\tt INT2}
type.

We show an actual SQL statement to create one of 
the indices;\codespc\\
{\footnotesize
\verb|CREATE INDEX twomass_xyzi_aaa0_i16xyz|\\
\verb|       ON Twomass_xyzi_aaa0|\\
\verb|          (fGetI16UVecI4(cxi,32400),|\\
\verb|           fGetI16UVecI4(cyi,32400),|\\
\verb|           fGetI16UVecI4(czi,32400));|\codespc\\
}%
where {\tt Twomass\_xyzi\_aaa0} is the name of a child table and 
{\tt fGetI16UVecI4(}{\it arg}{\tt ,32400)} scales and rounds the unit
vector of {\tt INT4} into that of {\tt INT2} having a range from
$-32400$ to $32400$.
This gives about $9''$ of spatial resolution in the worst case;
however, it does not cause any problems for typical searches.

The data size of all indices on table sets {\tt twomass\_xyzi} is about
10 Gbyte, which is small enough to be stored in a RAM disk and enables faster
file access.

\subsubsection{Stored Function}
\label{2mass_stored_function}

PostgreSQL has a constraint-exclusion (CE) feature to allow us a
seamless access to the partitioned tables.
If {\tt CHECK} constraints are included in the definitions of child tables,
the server parses an SQL statement referring a parent and accesses only
necessary child table(s).
We find that CE can improve the performance of a general one-time search.%
\footnt{We apply CE for rectangular search.  See \S\ref{rectangular}.} 
However, it is still not enough for repeating a radial search many times
within a small period of time.
It is desired that cross-identification can be also performed only with
an SQL statement that runs multiple radial searches.

To improve the performance further,
we create a stored function to access necessary child tables and perform
a radial search.
Although it is best to write the code in C from a performance
point of view,
an SQL execution in a C stored function is not supported in PostgreSQL.
Alternatively, PostgreSQL offers dynamic SQL execution in PL/pgSQL%
\footnt{PL/pgSQL is a procedural language for 
the PostgreSQL database system.}; 
therefore, we create stored functions in both PL/pgSQL and C.
We show a code to create a PL/pgSQL function that obtains an {\tt objID}
of the object nearest to the given position within the search region
(this function is used for cross-identification):\codespc\\
{\footnotesize
\verb|CREATE FUNCTION fTwomassGetNearestObjIDEq(|\\
\verb|           arg1 FLOAT8, arg2 FLOAT8, arg3 FLOAT8)|\\
\verb| RETURNS INT4|\\
\verb| AS $$|\\
\verb|  DECLARE|\\
\verb|   rt INT4;|\\
\verb|  BEGIN|\\
\verb|   EXECUTE _fTwomassGetSqlForRadialSearch(|\\
\verb|                    arg1, arg2, arg3, 32400, TRUE|\\
\verb|           ) INTO rt;|\\
\verb|   RETURN rt;|\\
\verb|  END|\\
\verb| $$ IMMUTABLE LANGUAGE 'plpgsql';|\codespc\\
}%
where {\tt \_fTwomassGetSqlForRadialSearch()} is a stored function
written in C that returns an SQL statement for radial search or
cross-identification (i.e., multiple radial searches), and
{\tt arg1}, {\tt arg2}, and {\tt arg3} are right ascension,
declination, and search radius, respectively.
The fourth and fifth arguments of {\tt \_fTwomassGetSqlForRadialSearch()}
are the scaling parameter of a composite expression index on 
({\tt cxi}, {\tt cyi}, {\tt czi}) and a switch 
to select either radial search ({\tt FALSE}) 
or cross-identification ({\tt TRUE}), respectively.

The function {\tt \_fTwomassGetSqlForRadialSearch()} knows range
information of declination for each child table of {\tt twomass\_xyzi} and
generates an appropriate SQL statement with the {\tt UNION ALL}%
\footnt{{\tt UNION ALL} is used 
to merge two results of {\tt SELECT} phrases.}
keyword (if required) to access necessary child table(s).
We show an example SQL statement generated by the function that searches
the nearest object from (0,0) in J2000 coordinates within 
a 10$'$ radius:\codespc\\
{\footnotesize
\verb|SELECT o.objid|\\
\verb|FROM|\\
\verb| (SELECT objid,|\\
\verb|         fDistanceArcMinXYZI4(cxi,cyi,czi,1,0,0)|\\
\verb|         AS distance|\\
\verb|  FROM twomass_xyzi_ace7|\\
\verb|  WHERE (fGetI16UVecI4(cxi,32400)|\\
\verb|         BETWEEN 32305 AND 32400) AND|\\
\verb|        (fGetI16UVecI4(cyi,32400)|\\
\verb|         BETWEEN -95 AND 95)      AND|\\
\verb|        (fGetI16UVecI4(czi,32400)|\\
\verb|         BETWEEN -95 AND 95)|\\
\verb|  UNION ALL|\\
\verb|  SELECT objid,|\\
\verb|         fDistanceArcMinXYZI4(cxi,cyi,czi,1,0,0)|\\
\verb|         AS distance|\\
\verb|  FROM twomass_xyzi_baa0|\\
\verb|  WHERE (fGetI16UVecI4(cxi,32400)|\\
\verb|         BETWEEN 32305 AND 32400) AND|\\
\verb|        (fGetI16UVecI4(cyi,32400)|\\
\verb|         BETWEEN -95 AND 95)      AND|\\
\verb|        (fGetI16UVecI4(czi,32400)|\\
\verb|         BETWEEN -95 AND 95)|\\
\verb| ) o|\\
\verb|WHERE o.distance <= 10|\\
\verb|ORDER BY o.distance|\\
\verb|LIMIT 1|\codespc\\
}%
Here {\tt fDistanceArcMinXYZI4()} is a stored function written in C to
obtain angular distance in arcminutes between two positions.

General radial search is performed by using the 
{\tt fTwomassGetNearbyObjEq()} function.
The source code of it is almost the same as that of 
{\tt fTwomassGetNearestObjIDEq()}.
See also the source files in our kit for details.

\subsubsection{Number of Partitions}
\label{partitioning}

\begin{figure}[!t]
 %\epsscale{1.0}
 %\plotone{compare.eps}
 \plotone{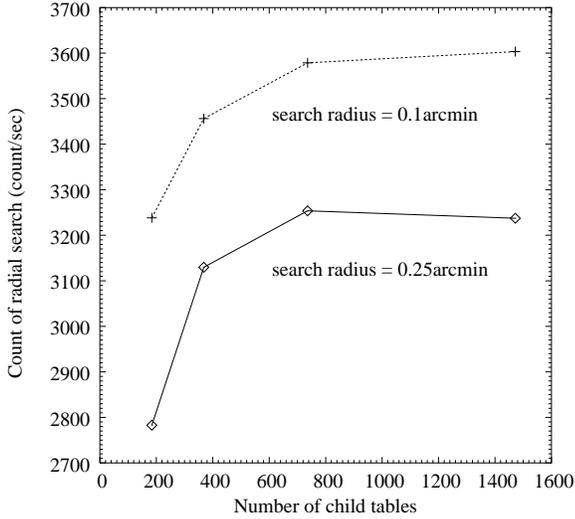}
 \caption{Performance of cross-identification between all objects
 of AKARI/IRC PSC and all objects of 2MASS PSC for different numbers of
 partitions of 2MASS PSC.
 The count of radial search per second is plotted against the number of
 child tables of 2MASS PSC.
 Hardware is a $2\times$ Opteron2384 CPU on a Supermicro dual-processor
 board with 64 Gbyte memory.
 }\label{fig:compare}
\end{figure}

The number of partitions of {\tt twomass\_xyzi} is an important factor for
performance.
A public version of 2MASS PSC is provided as 92 files divided by 
declination.%
\footnt{Strictly speaking, neighboring two files
have some overlapped area in declination.
We have to be careful in the area, when selecting objects
with strict declination ranges.
}
Therefore, we create 92$\times$$n$ partitions and
test the performance of cross-identification as multiple radial searches
for $n=2$, $n=4$, $n=8$, and $n=16$.
Figure \ref{fig:compare} is the result of the test.
All measurements were made with enough cached data in the main memory.
The case of $n=8$ shows significant improvement
compared with $n=2$ or $n=4$.
However, the $n=16$ case of radius = 0.25$'$
exhibits a slight decrease of the performance.
We suspect that increasing the {\tt UNION ALL} phrase in $n=16$ case causes the
slowdown and that around $n=8$ might be the best number of
partitions.

As a result, we apply $n=8$ for our system, which achieves comparable
performance with the case of 5 million objects in Figure \ref{fig:limit}.

\subsection{Performance with a Standard PC}

We test the performance of a radial search of 2MASS PSC using a standard
personal computer.
Table \ref{table:result_onesearch_2mass} is the test results, including
four cases of different search radii, and shows satisfactory
performance.
This test is made in the ``psql'' interactive terminal with the following SQL
statement:\codespc\\
{\tt \small
 SELECT count(*) \\ 
 FROM fTwomassGetNearbyObjEq(0, 0, {\it radius});
}\codespc\\
after the \verb|\timing| command.
All measurements were made with enough cached data in the main memory;
therefore, the actual search speed may be slower than them, due to the
bottleneck of disk I/O.
This bottleneck is generally reduced by performing many searches for a
long span that increases memory cache efficiency.

In actual searches,
we use this stored function and natural join between returned result and
{\tt twomass} table:\codespc\\
{\footnotesize
\verb|SELECT o.ra,o.dec,o.j_m,o.h_m,o.k_m,|\\
\verb|       o.j_msigcom,o.h_msigcom,o.k_msigcom,|\\
\verb|       n.distance|\\
\verb|FROM fTwomassGetNearbyObjEq(0,0,3) n, twomass o|\\
\verb|WHERE n.objid = o.objid;|\codespc\\
}%
where a join on the primary key
{\tt n.objid = o.objid} works fast enough.
See the document of our software kit for details.

\begin{table}[!t]
\begin{center}
\caption{The performance of radial search and rectangular search of
 2MASS PSC.
 Hardware is a Core2Quad Q9650 (3.0GHz) CPU on
 GIGABYTE GA-EX38-DS4 with 8 Gbyte DDR2-800 memory.
}\label{table:result_onesearch_2mass} {
%\tabcolsep=0.22zw
\vspace*{6.0pt}
\begin{tabular}{lcc}
\hline
\hline
Search criteria & No. Obj & Elapsed time \\
\hline
Radial ($r$$=$$1'$) & 2 & 0.001 s \\
Radial ($r$$=$$60'$) & 5198 & 0.022 s \\
Radial ($r$$=$$180'$) & 47632 & 0.149 s \\
Radial ($r$$=$$360'$) & 189784 & 0.484 s \\
Rect. ($2^\circ$$\times$$2^\circ$) & 6644 & 0.025 s \\
Rect. ($10^\circ$$\times$$10^\circ$) & 167266 & 0.386 s \\
\hline
\end{tabular} 
}
\end{center}
\end{table}

\section{Technical Design of Rectangular Search}
\label{rectangular}

As shown in Figure \ref{fig:relation}, 
we implement rectangular search using an approach similar to that of 
radial search.
Rectangular search is used as a one-time search
in major cases; therefore, we implement it so that we can obtain better
performance with minimum cost.
Although we introduced the CE feature of PostgreSQL in 
\S\ref{overview_sd} and \S\ref{2mass_stored_function},
we did not use it for radial search.
On the other hand,
we found that CE is suitable for rectangular search, and it simplifies
our implementation.

We create 92 partitions (child tables) for rectangular search
following the recommendation (less than 100 partitions) of the official
document of PostgreSQL, and we distribute all rows into child tables 
divided by their declination.
Then we create indices on all the child tables.
Here is an example to create one pair of indices:\codespc\\
{\footnotesize
\verb|CREATE INDEX twomass_j2000i_aaa_radeci|\\
\verb|       ON Twomass_j2000i_aaa (rai,deci);|\\
\verb|CREATE INDEX twomass_j2000i_aaa_decrai|\\
\verb|       ON Twomass_j2000i_aaa (deci,rai);|\codespc\\
}%
where {\tt Twomass\_j2000i\_aaa} is the name of a child table.
After creating a stored function for rectangular search in PL/pgSQL,
we can run a fast rectangular search like this:\codespc\\
{\footnotesize
\verb|SELECT o.ra,o.dec,o.j_m,o.h_m,o.k_m,|\\
\verb|       o.j_msigcom,o.h_msigcom,o.k_msigcom|\\
\verb|FROM fTwomassGetObjFromRectEq(0,0.1, 1,1.1) n,|\\
\verb|     twomass o|\\
\verb|WHERE n.objid = o.objid;|\codespc\\
}%
Note that we have to write the code of the stored function
{\tt fTwomassGetObjFromRectEq()} in PL/pgSQL,
since a search must be performed as dynamic SQL execution.
A stored function written in SQL only supports static SQL execution, under
which CE does not work.

Table \ref{table:result_onesearch_2mass} includes the performance of
two cases of rectangular search.
On average, the search speed with CE and partitioning is faster by more
than 10 times compared with the searches using an index on a single table.

Our stored function for rectangular search has several other minor
contrivances.
See the document and source files in the 2MASS Kit for details.

\section{Summary}
\label{summary}

We develop a software kit to construct a high-performance
astronomical catalog database 
supporting 2MASS PSC and several all-sky catalogs
on a standard personal computer.
The kit has a document that includes step-by-step instructions for installation
and a tutorial for database beginners,
and it utilizes open-source RDBMS. Therefore, 
any users can easily build their own catalog server 
without software licensing fees
and can search the catalogs with various criteria using SQL.

Out kit is tuned for optimal performance of
positional search, i.e., radial search and rectangular search.
We use an orthogonal coordinate system for database index to implement the
radial search that is most important in the positional search. 
This $x$$y$$z$-based method needs neither special processing for polar
singularity nor spatial splitting such as HTM or HEALPix.
Therefore, we can develop cost-effective astronomical database systems.

The implementation of radial search for relatively small (less than
several million entries) catalogs can be very simple, and good
performance is realized.
We also show that such simple implementation is not enough for the severe
search requirements of huge catalogs,
and we need additional techniques. 
We examine our revised implementation of radial search of the 2MASS PSC 
using techniques of table partitioning, composite expression index,
stored functions, etc.
Our performance tests of cross-identification of 
AKARI/IRC PSC (870,973 objects)
with 2MASS PSC (470,992,970 objects) 
achieve about a 2000 counts/s radial search using
a dual-processor server.
Additional tests using a standard personal computer also show
satisfactory performance of radial search with some typical search
radii.

We also present our simple implementation of fast rectangular search
for which the constraint-exclusion (CE) feature of PostgreSQL works
effectively to improve performance.

Commercial RDBMS products have often been used for services for huge
catalogs.
Our report shows that an open-source RDBMS product is also a good
choice to develop astronomical database services.

\acknowledgments

We acknowledge 
Satoshi Takita, Shinki Oyabu, Norio Ikeda and Yoshifusa Ita
for their test reports and suggestions about documentation.
We are grateful to Aniruddha R. Thakar, Yanxia Zhang
and Nigel Hambly for
giving us valuable information of cross-identification techniques.
We thank Kanoa Withington for reporting their use of 2MASS Kit at
Canada-France-Hawaii Telescope (CFHT) observatory.
We thank Shin-ichi Ichikawa and Masafumi Yagi for insightful
comments about application and literature of the orthogonal coordinate
system.
We acknowledge Issei Yamamura for his careful reading of the text.
We thank Jun Kuwamura for informing us about some technical tips of
PostgreSQL.
We thank the anonymous referee for his installation report and 
useful comments regarding this article.
We thank Japan PostgreSQL Users Group
for giving us the chance to talk about the 2MASS Kit
at the PostgreSQL Conference 2011 in Japan.
We thank Linux, PostgreSQL and other UNIX-related communities
for the development of various useful software.

\section*{Appendix A: Stored Functions in \S4.2}

Here, we explain each stored function 
in the definition of the {\tt fAkariIrcGetNearestObjIDEq()} function
in \S4.2.
All stored functions shown next are written in C.
\begin{itemize}
 \item
      {\tt fEq2X(ra,dec)}, {\tt fEq2Y(ra,dec)} and {\tt fEq2Z\\(ra,dec)}
      convert J2000 ({\tt ra},{\tt dec}) to 
      ({\tt cx}, {\tt cy}, {\tt cz}) of the unit vectors, respectively.
 \item
      {\tt fDistanceArcMinXYZ(cx1,\hspace{2pt}cy1,\hspace{2pt}cz1,\hspace{2pt}cx2,\\cy2,cz2)}
      returns the angular distance (in arcminutes) between two positions,
      ({\tt cx1},{\tt cy1},{\tt cz1}) and ({\tt cx2},{\tt cy2},{\tt cz2}).
 \item
      {\tt fArcMin2Rad(distance)}
      converts the distance in arcminutes to radians.
\end{itemize}

\section*{Appendix B: Performance Tuning}

Throughout the tests of cross-identification presented in this article,
we use PostgreSQL 8.4.5 with CentOS 5.5 64-bit on x86\_64 
compatible hardware.
We have adjusted the following points to obtain the highest performance for
our hardware:
\begin{itemize}
 \item
      Dynamic clocking of the CPU and others are disabled.
      We stop {\tt cpuspeed} using the {\tt chkconfig} command in OS and
      turn off C1E (Enhanced Halt) in BIOS.
      If they are enabled, the performance may decrease by more than
      10 \%.
 \item
      We set the {\tt readahead} parameter to 1024 using the {\tt hdparm} command.
      This sometimes improves by several percent compared with 
      the default value.
 \item
      We set the {\tt noatime} option of the {\tt mount} command for database storage.
\end{itemize}

\section*{Appendix C: Cross-Identification using SSD and Multicore CPU}

\begin{table}[!t]
\begin{center}
\caption{Results of match-up of AKARI/IRC PSC with 
2MASS PSC using multiple radial searches with multiple sessions.
Hardware is $2\times$ XEON X5650 (2.67GHz six-core 12-thread) CPU 
on a Supermicro dual-processor board and LSI 9211-4i HBA with a Crucial C300 SSD.
}\label{table:crossid_multicore} {
%\tabcolsep=0.4zw
\vspace*{6.0pt}
\begin{tabular}{ccc}
\hline
\hline
Sessions & Elapsed Time & N of Radial Search \\
\hline
1 & 7.72 minutes & 1881 counts/s \\
2 & 4.55 minutes & 3189 counts/s \\
4 & 2.50 minutes & 5796 counts/s \\
8 & 1.50 minutes & 9697 counts/s  \\
10 & 1.35 minutes & 10741 counts/s \\
12 & 1.28 minutes & 11333 counts/s \\
18 & 1.22 minutes & 11902 counts/s \\
24 & 1.22 minutes & 11898 counts/s \\
\hline
\end{tabular} 
}
\end{center}
\end{table}

One of the best methods to perform cross-identification with huge
catalogs is the plane sweep techniques 
\citep[][N. Hambly 2011, private communication]{dev05}.
Although 2MASS Kit supports cross-identification as multiple radial
searches, such a RDBMS-based approach might generally be unsuitable for
the best performance.
However, we found that satisfactory performance can be obtained using
2MASS Kit with recent inexpensive and high-speed SSDs and multicore CPUs.

We store the PostgreSQL database files of the index set and table set for
radial search (about 30 Gbyte) into a Crucial C300 MLC SSD,
and we test the performance of cross-identification between
AKARI/IRC PSC and 2MASS PSC using six-core (12-thread) CPUs on
a dual-processor board.
We connect multiple sessions to the PostgreSQL server
and execute the following SQL statements simultaneously:\codespc\\
{\footnotesize
{\tt SELECT count(}\\
\verb|     |{\tt fTwomassGetNearestObjIDEq(o.ra, o.dec, 0.25))}\\
{\tt FROM ( SELECT * FROM AkariIrc }\\
\verb|       |{\tt WHERE objid \% $n$ = $m$ ORDER BY dec ) o;}}\codespc\\
where $n$ is the number of sessions, and a unique sequential number
beginning with $0$ is assigned to $m$ for each session.
For example, we set $n=4$ and $m=0,1,2$ and $3$ for cross-identification
using four sessions (threads).

The results are shown in Table \ref{table:crossid_multicore}.
Less than 10 sessions show significant scaling factor.
Recent mainstream CPUs have four cores or more; therefore,
RDBMS-based multiple radial searches might become a good choice for
several situations.

\label{lastpage}

\end{document}